\documentclass[proof]{WileyASNA-v1}

\articletype{Article Type}%

\received{26 April 2016}
\revised{6 June 2016}
\accepted{6 June 2016}

\begin{document}

\title{Fourier spectral-timing techniques for the study of accreting black holes}

\author[1]{Adam Ingram*}

\address[1]{\orgdiv{School of Mathematics, Statistics and Physics}, \orgname{Newcastle University}, \orgaddress{Herschel Building, Newcastle Upon Tyne, \country{UK}}}

\corres{*\email{adam.ingram@newcastle.ac.uk}}


\abstract{The X-ray signal from active galactic nuclei and black hole X-ray binaries is highly variable on a range of timescales. This variability can be exploited to map the region of interest close to the black hole, which is far too small to directly image for all but two black holes in the Universe. Spectral-timing techniques provide causal information by combining timing and spectral information. I present a brief review of such techniques, focusing on two examples: X-ray reverberation mapping and phase-resolved spectroscopy of low frequency quasi-periodic oscillations (LF QPOs). The former provides a means to diagnose the accretion geometry and measure parameters such as black hole mass, and the latter gives perhaps the best constraints we currently have as to the enigmatic LF QPO mechanism.}


\jnlcitation{\cname{%
\author{A. Ingram}} (\cyear{2022}), 
\ctitle{Fourier spectral-timing techniques for the study of accreting black holes}, \cjournal{Astronomical Notes}, \cvol{2022;00:1--6}.}


\maketitle

\section{Introduction}
\label{sec:intro}

Accretion onto black holes (BHs) can efficiently produce huge X-ray luminosities, lighting up the vicinity of the horizon. This process happens in active galactic nuclei (AGNs), whereby a supermassive BH accretes material from its host galaxy, and in BH X-ray binaries (XRBs), whereby a stellar-mass BH accretes from its binary partner. In both object classes, accretion occurs via a thin, accretion disk and, close to the hole, a cloud of hot electrons commonly referred to as the corona. The shape and physical nature of the corona is still a subject of hot debate, with models including a layer sandwiching the disk \citep{Galeev1979,Haardt1991}, a large scale-height accretion flow located inside of a truncated disk \citep{Eardley1975,Ichimaru1977} and a compact or vertically extended region at the base of the out-flowing jet \citep{Miyamoto1991,Markoff2005,Miniutti2004}. The region of interest is far too small to directly image, necessitating sophisticated mapping techniques to infer the nature of the corona. Spectral-timing methods do just this, exploiting the rapid time variability routinely observed in the X-ray flux. Since signal to noise is typically low for each variability cycle but we observe many cycles, such rapid variability is best studied in the Fourier domain.

The X-ray spectrum includes a multi-temperature blackbody component contributed by the disk \citep{Shakura1973,Novikov1973} and a cut-off power-law component contributed by Compton up-scattering in the corona \citep{Thorne1975,Sunyaev1979}. X-rays from the corona that irradiate the disk are reprocessed and re-emitted by the disk to form a so-called reflection spectrum that includes features such as a $\sim 6.4$ keV iron K$\alpha$ line and a broad `Compton hump' peaking at $\sim 20-30$ keV \citep{Matt1991,Ross2005,Garcia2013}. General relativistic effects distort the reflection spectrum seen by the observer, leading to the iron line being broadened by rapid disk orbital motion and skewed by gravitational redshift and Doppler boosting \citep{Fabian1989}. The spectrum of XRBs is observed to transition from the power-law dominated hard state to the disk dominated soft state, via the intermediate state, on timescales of $\sim$weeks-years \citep[e.g.][]{Belloni2010,Done2007}. AGN, on the other hand, evolve much more slowly, since all characteristic timescales scale linearly with BH mass.

\begin{figure*}
\vspace{-0.5cm}
\begin{centering}
\includegraphics[width=17.9cm,trim=0.0cm 8.5cm 0.0cm 0.0cm,clip=true]{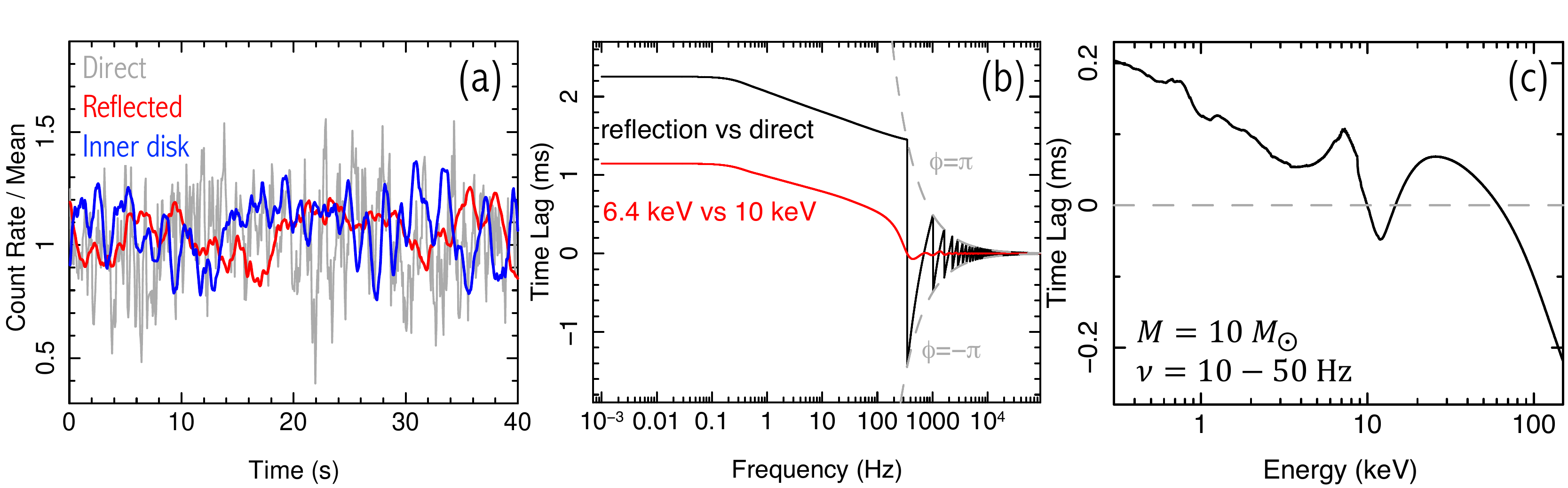}
\end{centering}
\vspace{-0.5cm}
\caption{Demonstrations of reverberation. a) A simulated direct signal (grey) smoothed and lagged (red and blue). b) Time lag vs Fourier frequency between direct and reflected signals (black) and between two energy bands (red). c) Lag-energy spectrum calculated with the same reverberation model as (b).}
\label{fig:reverb}       
\end{figure*}

The rapid X-ray variability includes broad band stochastic noise and quasi-periodic oscillations (QPOs). The broad band noise has a roughly constant root-mean-squared (rms) variability amplitude between low and high frequency breaks that scale with spectral state and BH mass \citep{McHardy2006}. Different kinds of QPO are observed in XRBs, with the main division being between low frequency (LF) QPOs (centroid frequency $\nu_{\rm qpo} \lesssim 30$ Hz) and high frequency (HF) QPOs ($\nu_{\rm qpo} \gtrsim 30$ Hz). Whereas HF QPOs are very weak and rare features \citep{Belloni2012}, LF QPOs are very strong and commonly observed \citep{Ingram2019b}. The X-ray timing properties evolve in a manner tightly correlated with the spectral state transitions \citep{Wijnands1999,Psaltis1999}. In AGN, QPOs are notoriously difficult to observe due to the long timescales and irregular sampling involved \citep{Vaughan2005}. As such, most known AGN QPOs are thought to be mass-scaled analogies of HF QPOs \citep[e.g.][]{Gierlinski2008}. 


In this proceeding of the conference `Black hole accretion under the X-ray microscope' (14-17 June 2022 at ESAC, Madrid), I will briefly review the spectral-timing methods at our disposal to learn more about the accretion geometry, the origin of QPOs and even attempt to measure the fundamental BH parameters: mass and spin. I will focus on two examples: in Section \ref{sec:reverb} X-ray reverberation mapping, and in Section \ref{sec:QPOs} phase-resolved spectroscopy of LF QPOs. Conclusions are presented in Section \ref{sec:conc}.

\section{Reverberation mapping}
\label{sec:reverb}

Reverberation mapping exploits the light crossing delay between X-rays that travel directly from the corona to the observer, and those that take a longer path by first reflecting off the accretion disk. This is demonstrated conceptually in Fig \ref{fig:reverb}a. The reflected signal (red) is a lagged and smoothed version of the direct signal (grey, generated using the algorithm of \citealt{Timmer1995}). The smoothing is due to the different travel time of rays that reflect off of different parts of the disk. Mathematically, this lagging and smoothing can be achieved by convolving with an \textit{impulse-response function} \citep[e.g.][]{Uttley2014,Bambi2021}. Here, I have simply used a top-hat function (with parameters designed to greatly exaggerate the salient effects), whereas modern reverberation models calculate an impulse-response function by assuming a geometry and accounting for all relativistic effects \citep[e.g.][]{Ingram2019}. Since the light crossing delays associated with reflection from the inner disk are shorter than those for the outer disk, the reflection signal from the inner disk only (blue) is lagged and smoothed to a lesser degree than that from the entire disk (red). Therefore, if we could isolate only the fastest variability, we could isolate the signal from the inner disk only to build up a map of the disk. How can we isolate different variability timescales? We take a Fourier transform!

Fig \ref{fig:reverb}b shows the time lag between the direct and reflected signal (black line) as a function of Fourier frequency, this time with the impulse-response function calculated for an XRB using the reverberation model \textsc{reltrans} \citep{Ingram2019}. The Fourier frequency-dependent phase lag between two signals can be calculated from the argument of the complex cross-spectrum \citep{vanderKlis1987} and the time lag is then recovered from division by $2\pi \nu$, where $\nu$ is Fourier frequency (see e.g. \citealt{Uttley2014} for a detailed review). The lags in Fig \ref{fig:reverb}b are independent of the driving signal, they are purely governed by the impulse-response function. The $\sim 2.2$ ms lag seen at low frequencies is the extra time it takes for most of the reflected flux to arrive at our detector relative to the direct flux. The lag reduces for higher frequencies ($\gtrsim 0.1$ Hz) because higher Fourier frequencies isolate reflection from smaller disk radii, for which the light crossing delay is shorter. For the highest frequencies, we see \textit{phase-wrapping}. This happens when the reverberation lag is greater than the Fourier period, or in other words the reverberation phase lag is $>\pi$ or $<-\pi$ radians. The same effect can be seen in movies when car wheels appear to be moving backwards because they almost complete a full rotation between camera frames.


Unfortunately, X-ray photons detected from accreting BHs do not come labelled as either direct or reflected. They must instead be separated out by their different spectra\footnote{In principle they can also be separated out by their different polarization properties but this is challenging with current instrumentation \citep{Ingram2022a}}: a cut-off power law continuum for the direct photons and a reflection spectrum for the reflected photons. The red line in Fig \ref{fig:reverb}b shows the time lag between $6.4$ keV and $10$ keV photons. The $6.4$ keV band (containing the iron line) lags behind the 10 keV band because it includes a greater fraction of reflected photons. However, the lag is \textit{diluted} compared to the true reverberation lag because each band includes some direct and some reflected photons.


\begin{figure}[b]
\vspace{-0.5cm}
\begin{centering}
\includegraphics[width=8.0cm,trim=21.0cm 0.0cm 0.0cm 0.5cm,clip=true]{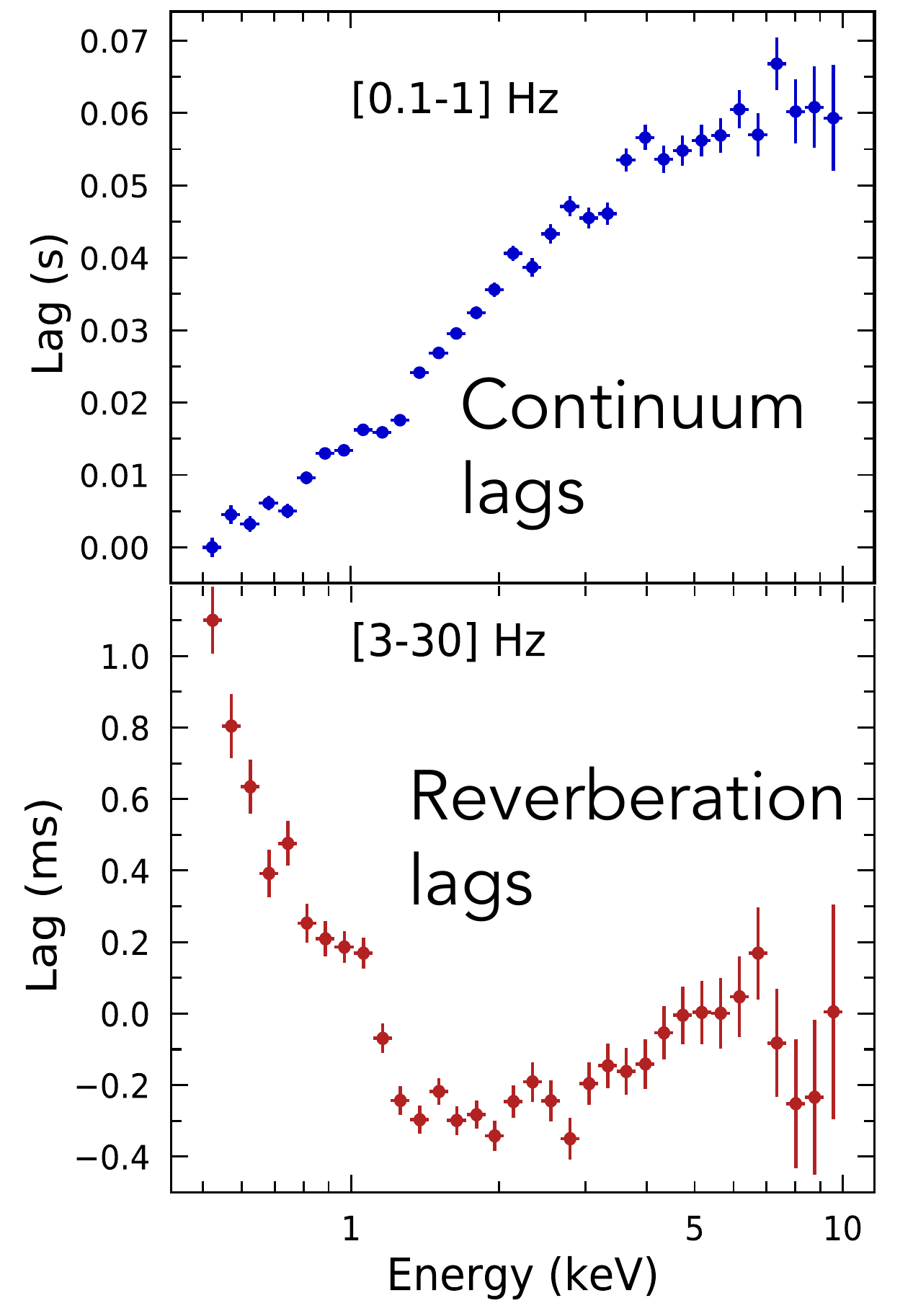}
\end{centering}
\vspace{0.0cm}
\caption{Lag-energy spectrum of MAXI J1820+070 demonstrating featureless continuum lags at low frequencies (top) and reverberation features at high frequencies (bottom).}
\label{fig:revobs}       
\end{figure}

The clearest observational manifestation of reverberation is in the \textit{lag-energy spectrum}, which is derived from a series of cross-spectra between the count rate in each energy band and one common reference band \citep{Kotov2001}. Fig \ref{fig:reverb}c shows by how much time photons of each energy lag behind 10 keV photons (i.e. here the reference band is narrow and centered on 10 keV, but in practice using the full telescope band-pass maximises signal to noise; e.g. \citealt{Ingram2019a}). We see an iron line and Compton hump feature in the lag-energy spectrum, which result from there being more reflection, and therefore less dilution, at these energies. So do we see such features in observational lag-energy spectra? The short answer is yes. Fig \ref{fig:revobs} (bottom) shows the clearest example of an iron K reverberation feature for an XRB \citep{Kara2019}. Similar iron K reverberation features have been observed for $\sim 25$ AGN \citep{Kara2016} and the amplitude of the lag scales with BH mass \citep{DeMarco2013}. Therefore, the lag is \textit{much} larger for AGN than for XRBs, simply because the light crossing time of the system scales linearly with BH mass.

However, reverberation lags are only visible for high Fourier frequencies ($\nu \gtrsim 3~[10~M_\odot]/M$ in practice; e.g. \citealt{Kara2016}), i.e. considering only the most rapid variability in the system.
The low frequency lag-energy spectrum is instead featureless (Fig \ref{fig:revobs} top). Since there are no reverberation features, these lags are not associated with reflection and must be caused by non-linear spectral variability of the continuum. Note that these `continuum lags' are much larger than the reverberation lags. It is very likely that reverberation lags are still present at low frequency, but they are simply drowned out by the much larger continuum lags. In fact, it is often possible to see a dip in the low frequency lag-energy spectrum at $\sim 6.4$ keV \citep{Kotov2001}. This can be explained by reverberation: the comparatively small reverberation lag dilutes the larger continuum lag more at the iron line energy than at any other, leading to a dip \citep{Mastroserio2018}.

The simplest way to proceed with reverberation modelling is to simply ignore the continuum lags by modelling only the high frequency lag-energy spectrum \citep{Cackett2014,Ingram2019}. Moreover, the simplest model to employ is the lamppost model, whereby the corona is assumed to be a point-like source some height $h$ above the BH \citep{Matt1992}. However, this approach provides only weak constraints on the BH mass, since the same reverberation lag results if $h$ is many small gravitational radii ($r_g=GM/c^2$) or only a few large gravitational radii. The degeneracy can be broken either by considering many observations, over which the coronal height changes but the BH mass stays the same \citep{Alston2020}, or by modelling multiple frequency ranges -- requiring a model for the continuum lags. 
The continuum lags are typically attributed to inward propagating fluctuations in mass accretion rate \citep{Lyubarskii1997,Arevalo2006,Ingram2013}, but accurately modelling this process and reverberation mapping simultaneously is too computationally intensive to fit the model directly to data \citep{Wilkins2016}. The `two blobs' model of \citet{Chainakun2017} simplifies the problem by considering propagating fluctuations between two lamppost sources. The model \textsc{reltrans} \citep{Ingram2019} instead accounts for continuum lags by introducing fluctuations in the continuum power-law index \citep{Mastroserio2018}.

\textsc{reltrans} has been used to fit the lag-energy spectrum in multiple frequency ranges, simultaneously with the time-averaged spectrum, for several AGNs and XRBs \citep[e.g.][]{Mastroserio2020,Wang2021}. For Cygnus X-1, it was even possible to also fit the \textit{rms-energy spectrum} (the energy dependence of rms variability amplitude) in multiple frequency ranges. This extra constraint informs on how destructive interference between rays that reflected from different parts of the disk washes out variability in the reflection spectrum (i.e. exactly the smoothing shown in Fig \ref{fig:reverb}a). The resulting mass estimate of $M = 26.0\pm 5.5~M_\odot$ \citep{Mastroserio2019} agrees with the dynamical measurement of $M = 21.2 \pm 2.2~M_\odot$ \citep{Miller-Jones2021}. Other objects have proved more challenging: the \cite{Mastroserio2020} X-ray reverberation mass of Mrk 335 disagrees with existing optical reverberation measurements \citep{Peterson2004}, and fitting the lag-energy spectra of MAXI J1820+070 yields a very different source height to fits of the same model to the time-averaged spectrum \citep{Wang2021}. Part of the problem may be the over-simplicity of the lamppost geometry. Also, the fits to Mrk 335 and MAXI J1820+070 (respectively using XMM and NICER data) may be mainly driven by the soft X-rays, which carry the most model uncertainty and are not present in the RXTE data used for the Cygnus X-1 mass estimate.

\section{Quasi-periodic oscillations}
\label{sec:QPOs}


Spectral-timing methods are also key tools to diagnose the mechanism behind QPOs. The best example to focus on for this review is that of Type-C QPOs ($0.1~{\rm Hz} \lesssim \nu_{\rm qpo} \lesssim 30$ Hz), which are by far the most commonly observed QPO class (see \citealt{Casella2005} for a description of the Type A-B-C classification) and are also very strong signals on which to apply spectral-timing methods. Studies of the QPO rms-energy spectrum reveal that the Comptonised component oscillates with higher amplitude than the disk \citep{Sobolewska2006}, indicating that the QPO is modulated in the corona. Population studies provide strong evidence that Type-C QPOs are geometrical in origin; i.e. the accretion geometry is oscillating quasi-periodically. Firstly, the QPO amplitude is larger for higher inclination (more edge-on) systems \citep{Motta2015}. Secondly, low inclination sources always show soft phase lags\footnote{Note that the cyclical nature of QPOs makes it more intuitive to think about phase lags, whereas reverberation is more intuitively considered in terms of time lags.} on the QPO fundamental frequency, whereas high inclination objects display soft lags for $\nu_{\rm qpo} \gtrsim 2$ Hz \citep{vandeneijnden2017}.

Currently the most powerful QPO diagnostic is provided by \textit{QPO phase-resolved spectroscopy}: constraining the flux-energy spectrum as a function of QPO phase. Of particular interest is the QPO phase-dependence of the iron line profile, since this can in principle be used to reconstruct the accretion geometry as a function of QPO phase. A prominent class of QPO models attribute the QPO to Lense-Thirring precession (see \citealt{Ingram2019b} for details) -- a nodal precession of orbits driven by the BH spin. In the model of \citet{Ingram2009} it is the corona that undergoes Lense-Thirring precession, in which case the iron line is predicted to rock from red to blue shifted throughout each QPO cycle as respectively receding and approaching disk azimuths are preferentially illuminated \citep{Ingram2012}. QPO phase-resolved spectroscopy is, however, technically challenging because QPO phase does not increase linearly, or even deterministically, with time \citep[e.g.][]{Morgan1997}. The QPO waveform can therefore not be measured by simply folding on the QPO period. One method to circumvent this problem is to first attempt to isolate the QPO signal from the broad band noise by filtering the light curve, and then assign QPO phase values to each time bin from peaks, troughs and mean crossings of the filtered signal \citep{Tomsick2001,vandeneijnden2016,Henric2016}.

\begin{figure}
\begin{centering}
\includegraphics[width=8.5cm,trim=21.0cm 0.0cm -1.0cm 0.0cm,clip=true]{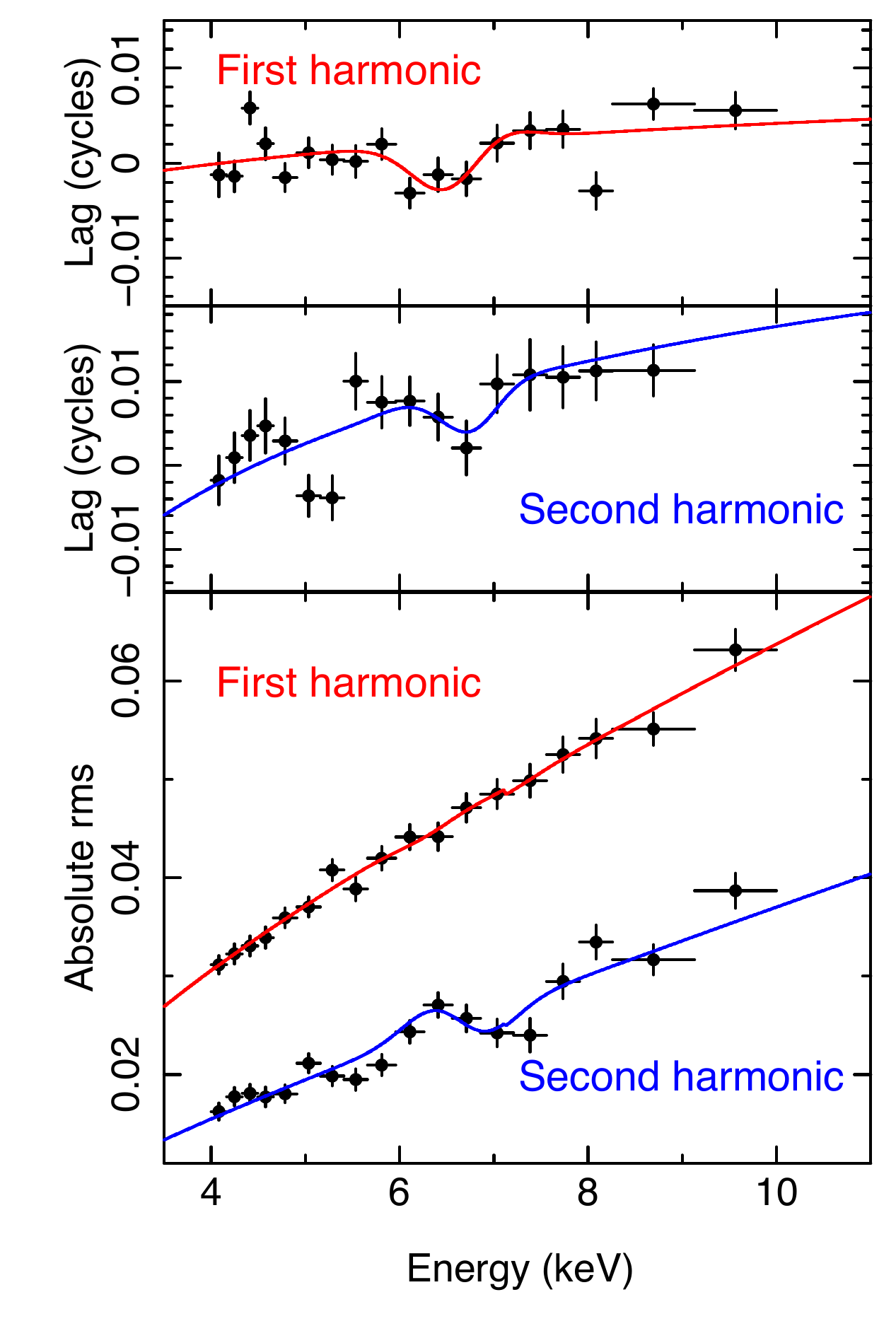}
\end{centering}
\vspace{0.0cm}
\caption{Lag-energy spectrum (top) and rms-energy spectrum (bottom) of the QPO fundamental (red) and second harmonic (blue). Solid lines are generated by a model in which a power law plus Gaussian iron line spectral model oscillates periodically on both harmonics. Black points are a $\sim 65$ ks exposure sub-set of XMM-Newton data from the $200$ ks XMM-Newton plus $70$ ks NuSTAR observation analysed by \cite{Ingram2016}.}
\label{fig:prs}       
\end{figure}

However, a higher signal to noise can be achieved by \textit{reconstructing} the phase-resolved spectrum from Fourier spectral-timing products. Almost all of the information is contained in the rms-energy spectrum and the lag-energy spectrum. Fig \ref{fig:prs} demonstrates this: lines are generated by a model in which a power-law continuum and Gaussian iron line oscillate periodically with two harmonics. Oscillations of the power-law index and normalisation drive the continuum shape of the lag-energy (top) and rms-energy (bottom) spectra at the first (red) and second (blue) harmonics. Oscillations of the Gaussian centroid, width and normalisation drive the `wiggles' around $\sim 6.4$ keV. The predicted quasi-periodic modulation of the iron line centroid energy therefore translates to predicted wiggles in the lag-energy and rms-energy spectra at the QPO first and second harmonic.

Since the first two harmonics typically dominate the QPO signal, the only information missing from the full phase-resolved spectra here is the phase difference between the two harmonics \citep{Ingram2015,DeRuiter2019}. This is because the rms-energy spectrum measures the relative amplitude of the harmonics and the lag-energy spectra measures how the phase of each harmonic in each energy band corresponds to the phase of the same harmonic in the reference band, but does not compare the phase of the two harmonics. The method of \cite{Ingram2015}, later improved upon by \cite{Ingram2016} and \cite{Nathan2022}, essentially includes this extra phase difference information by adjusting the second harmonic lag-energy spectra by the phase difference measured in the reference band. On the other hand, the cross-correlation method of \cite{Stevens2016} ignores the phase difference information. 

\cite{Ingram2016} used the Fourier reconstruction method to discover a QPO phase-dependence of the iron line centroid energy in H 1743-322 (black points in Fig \ref{fig:prs}). This constitutes strong evidence in favour of a precession origin of the Type-C QPO. However, other models can potentially explain this, for example a spiral density wave in the disk \citep{Tagger1999}. More advanced phase-resolved spectral modelling of the same data revealed a QPO phase-dependence of the reflection fraction \citep{Ingram2017}, which provides definitive evidence for a geometric QPO origin. Similar results were obtained for GRS 1915+105 by \cite{Nathan2022}. A detailed review of QPO models and observational constraints can be found in \cite{Ingram2019b}.

\section{Conclusions}
\label{sec:conc}

Spectral-timing techniques provide insights into the inner workings of accreting BHs by providing causal information; for example whether variations in one spectral component precede or delay those in another. The key tool for spectral-timing is the Fourier cross-spectrum, since the variability amplitude can be derived from its modulus and the time lag between two signals can be derived from its argument. This short review has covered two prominent examples of Fourier spectral-timing: X-ray reverberation mapping and QPO phase-resolved spectroscopy; both of which make heavy use of the cross-spectrum. I have briefly reviewed the key concepts that underpin both methods, and summarised the current state of the art of the field. Both methods with benefit greatly from future high throughput X-ray missions, since higher count rates reduce counting errors. In particular, this will enable reverberation mapping studies to push to higher Fourier frequencies (ideally into the phase-wrapping regime) and will enable the phase-resolved spectroscopic studies we currently conduct for LF QPOs to also be applied to HF QPOs. Combination with X-ray polarisation measurements, which are now finally achievable \citep{Henric2022} and in the near future may be available with greatly improved signal to nose \citep{DeRosa2019}, will hugely complement the spectral-timing tools discussed here.

\bibliography{biblio}%



\end{document}